# Exploring the properties of lead-free hybrid double perovskites using a combined computational-experimental approach

Zeyu Deng, Fengxia Wei, Shijing Sun, Gregor Kieslich, Anthony K. Cheetham* and Paul D. Bristowe*

Department of Materials Science and Metallurgy, University of Cambridge, 27 Charles Babbage Road, Cambridge CB3 0FS UK

*Corresponding Author: akc30@cam.ac.uk, pdb1000@cam.ac.uk

**Abstract**

**Density functional theory screening of the hybrid double perovskites $(MA)_2B^IBiX_6$ ($B^I$=K,Cu,Ag,Tl; X=Cl,Br,I) shows that systems with band gaps similar to those of the $MAPbX_3$ lead compounds can be expected for $B^I$=Cu,Ag,Tl. Motivated by these findings, $(MA)_2TlBiBr_6$, isoelectronic with $MAPbBr_3$, was synthesised and found to have a band gap of ~2.0eV.**

The remarkable performance of hybrid perovskite-based solar cells has launched a new paradigm in the area of photovoltaic research.[1] Facilitated by the processing flexibility of hybrid lead halide perovskites, *e.g.* $[CH_3NH_3]PbI_3$ and $[(NH_2)_2CH]PbI_3$, the turn-over efficiencies have soared to over 20% within a short period of time.[2–4] In materials such as $[CH_3NH_3]PbI_3$, which is the most widely studied phase, a 3D inorganic framework of the ReO$_3$-type is formed by $[PbI_3]^-$, whilst the $[CH_3NH_3]^+$ cation fills the dodecahedral void to form a perovskite-like structure. However, the inorganic $[PbX_3]^-$ framework with X = Cl$^-$, Br$^-$ and I$^-$ exhibits a highly dynamic character which is apparent in the thermal ellipsoids of the halide anions obtained from single crystal X-ray diffraction.[5,6] Together with the motion of the protonated cation, these materials exhibit disorder that poses challenges to crystallographers and computational scientists alike. For instance, expensive molecular dynamics (MD) simulations are usually required to access the electronic structures of the cubic and tetragonal polymorphs of $[CH_3NH_3]PbI_3$.[7,8] The exceptional performance of hybrid perovskites as light absorbing materials is attributed to the nearly perfect intrinsic properties, such as tuneable band-gaps, long carrier lifetimes and fairly small effective masses, amongst other factors.[9,10] However, the moisture sensitivity and toxicity of lead may limit wider commercialisation of such materials.[11] These issues represent the biggest challenges in the field to date and are topics of great current interest.

In the search for lead-free alternatives, Bi$^{3+}$ based compounds have recently attracted attention. Examples include the preparation of low-dimensional compounds such as $A_3Bi_2I_9$ (A = K$^+$, Cs$^+$, Rb$^+$, $[NH_4]^+$ and $[CH_3NH_3]^+$), as well as the investigation of (hybrid) double perovskites.[12–14] In particular, the study of double perovskites in which the 3D perovskite-motif is maintained seems to be a promising avenue. In such materials, the perovskite formula is essentially doubled and the divalent metal cations, *e.g.* Pb$^{2+}$ cations, are replaced by a monovalent and a trivalent cation, *e.g.* Na$^+$ and Bi$^{3+}$. The inorganic phases of general formula $A_2B^IB^{III}X_6$, with A = Cs$^+$, $B^I$ = Na$^+$, Ag$^+$, $B^{III}$ = Bi$^{3+}$ and X = Cl$^-$, Br$^-$, have been investigated since the 70s,[15–18] initially in the context of ferroelectrics. Recently, we reported the synthesis and properties of the first inorganic-organic *hybrid* double perovskite $(MA)_2KBiCl_6$ (MA$^+$ = $[CH_3NH_3]^+$).[19] At this point no double perovskite iodides have been reported. The application of these systems as photovoltaic absorbers is limited due to unfavourable electronic properties, originating from the ionic character of Cl$^-$ and Br$^-$ bonds. For example, the band-gap of our recent $(MA)_2KBiCl_6$ is approximately 3 eV. However, it would be expected that the use of Pearson-softer anions and cations, such as Br$^-$ or I$^-$ with, say, Tl$^+$ or Ag$^+$ as $B^I$ may lead to enhanced optoelectronic properties. Motivated by our previous findings, we now report an in-depth computational screening study on lead-free hybrid double perovskites as potential photovoltaic absorbers. In particular, we apply density function theory (DFT) calculations to study the series of hybrid double perovskites $(MA)_2B^IB^{III}X_6$ with $B^I$ = K$^+$, Tl$^+$, Ag$^+$, Cu$^+$, $B^{III}$ = Bi$^{3+}$, and X = Cl$^-$, Br$^-$ and I$^-$ (Figure 1). Additionally, we describe the synthesis and optical properties of a new double hybrid perovskite, $(MA)_2TlBiBr_6$, which is isoelectronic with MAPbBr$_3$ and has a much narrower band-gap than $(MA)_2KBiCl_6$.

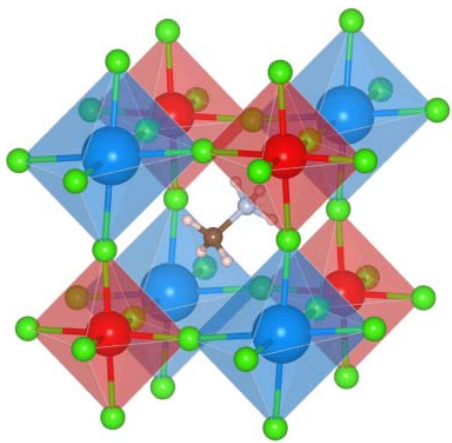

Figure 1 Crystal structure of $(MA)_2B^IB^{III}X_6$. Blue: $B^I$, red: $B^{III}$, pink: H, brown: C, grey: N, green: X. $B^IX_6$ and $B^{III}X_6$ octahedra are shown in blue and red, respectively. Structure was produced using the VESTA program.[20]

The DFT calculations were performed using the *Vienna Ab Initio Simulation Package* (VASP).[21,22] Details of the methodology, including pseudopotentials, exchange-correlation functional, k-point mesh, kinetic energy cut-off and convergence criteria, are given in the SI. Importantly for hybrid materials containing heavy elements, van der Waals forces and spin-orbit coupling (SOC) were included. Since $(MA)_2KBiCl_6$ is observed to crystallise in the $R\bar{3}m$ space group, the same symmetry was applied to the calculated structures both initially and during geometry optimisation. This allows for direct comparison with the experiments and enables trends in behaviour to be studied. In addition to the group of 12 double hybrid perovskites listed above, a further 3 structures of the form $MAPbX_3$ with $B^I = B^{III} = Pb$ were computed as a reference set. Following equilibration, the atomic and electronic structures of each perovskite were analysed with particular focus on the effect of the ionic radius on the band gap and the mechanical stability of the structures.

The computed lattice constants (referred to hexagonal axes), equilibrium volumes and $c/a$ ratios of the 15 structures as a function of the X anion radii are given in Table S1 and displayed in Figure S1. It is seen that the first two quantities increase with increasing anion radius, as expected ($r_{Cl}$ = 180 pm, $r_{Br}$ = 196 pm and $r_I$ = 220 pm).[23] The almost constant variation in c/a ratio indicates that the lattice expansion is approximately isotropic. The various interatomic bond distances (e.g. $B^I$-X, N⋯X and C⋯X) as a function of the X radii are given in Table S2 and Figure S2, and again show an increasing trend with anion radius. Furthermore, as the effective radius of the $B^I$ cation increases the bond distances also generally increase for each halide ($r_{Cu}$ = 77 pm, $r_{Ag}$ = 115 pm, $r_K$ = 138 pm and $r_{Tl}$ = 150 pm).[23] Each of these results is consistent with expectations based on the relative sizes of the ions. Hydrogen bonding contributes to the stability of each structure and it can be seen from Figure S2 that X⋯H bond distances, for example, increase as the X anion becomes less electronegative, indicating a weakening of the H-bonds. The various bond angles (e.g. $B^I$-X-$B^{III}$, X-$B^{III}$-X and C-H⋯X) as a function of the X radii are given in Table S3 and Figure S3 and do not show a strong variation with anion radius. In addition, the angles do not deviate much from those expected for the ideal geometry, e.g. 180° for $B^I$-X-$B^{III}$, indicating that tilting of the $B^IX_6$ and $B^{III}X_6$ octahedra is relatively small. For example, Pb-I-Pb is 177.5° in $MAPbI_3$ whereas in the low temperature, orthorhombic $MAPbI_3$, which is known to exhibit significant tilting, the equivalent computed angle is 145°.[24]

Figure 2 and Table S4 give the calculated band gaps of the 15 structures as a function of the X anion radii, where it is seen that the gap decreases by up to 1 eV as the X radii increase. Furthermore, the $B^I$ cation can significantly lower the band gap for a given halide. In all cases the largest gap is found when $B^I$ = K (e.g. 3.02 eV for $(MA)_2KBiCl_6$) and the smallest gap when $B^I$ = Cu (e.g. 0.28 eV for $(MA)_2CuBiI_6$). The calculated gaps for the $MAPbX_3$ reference structures are close to previous calculations for their cubic counterparts.[25] Figure S4 shows that the band gap systematically decreases as $B^I$-X-$B^{III}$ approaches 180°, which is in agreement with previous calculations on single perovskites, showing that smaller octahedral tilts result in reduced band gaps.[26]

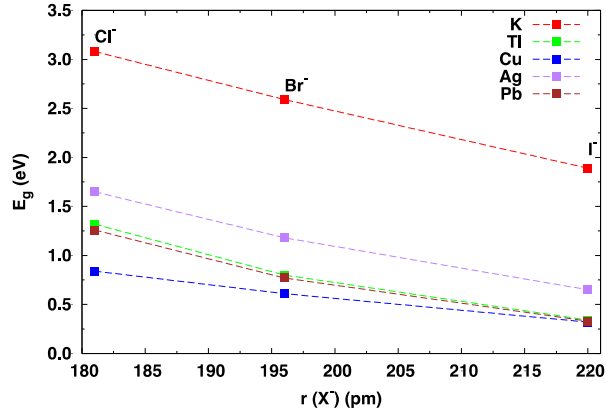

Figure 2 Computed band gaps of (MA)$_2$B$^I$B$^{III}$X$_6$ as a function of X anion radius, including the MAPbX$_3$ reference structures.

To understand the origin of these effects, the projected densities of states (PDOS) were calculated and are shown in Figures S5-S7, where attention is paid to the band edges. First, it is seen that the MA$^+$ cation does not contribute to states at the band edges, in agreement with previous studies on MAPbI$_3$.[27] It is interesting to note that if the MA$^+$ cation is replaced with Cs$^+$ in the double perovskites Cs$_2$AgBiCl$_6$ and Cs$_2$AgBiBr$_6$, the Cs$^+$ cation also does not contribute to states near the band edges.[17] Second, X-p states make a large contribution at both the valence band maximum (VBM) and the conduction band minimum (CBM). This indicates that choosing the appropriate halogen, specifically iodine, is important for obtaining a small band gap, semiconducting material. Similarly, appropriate choice of the B$^I$ cation is essential because using K$^+$, for example, which is strongly ionic, leads to an undesirably large band gap. This is because only X-p, Bi-6s and Bi-6p orbitals contribute to the band edge states (Figure S5). However, using Tl$^+$, Tl-6s and Tl-6p orbitals also make a contribution, which reduces the band gap significantly (Figure S5). This is similar to the situation in MAPbX$_3$ where Pb-6s and Pb-6p orbitals are present near the band edges (Figure S7). Incorporating Cu$^+$ onto the B$^I$ site further reduces the band gap due to the presence of Cu-3d orbitals near the VBM (Figure S6).

The electronic band structures of the 15 perovskites are collected in Figure S8. The band shapes near the band edges, which determine the carrier effective masses, fluctuate more with respect to the X anion than with respect to the B$^I$ cation. B$^I$ = K$^+$ produces the flattest bands while B$^I$ = Tl$^+$ or Pb$^+$ gives the most curved bands. Although all the structures contain the MA$^+$ cation, their band gaps and carrier effective masses differ widely. The location of the VBM and CBM within the Brillouin Zone depend mostly on the B$^I$ cation, except for K$^+$, and are largely independent of the X anion, as shown in Table S5. When B$^I$= Tl$^+$ the band gap is direct, but when B$^I$ = Cu$^+$ or Ag$^+$ it is indirect due to the large density of 3d states near the VBM of the latter. It is seen that the band structures of (MA)$_2$CuBiX$_6$ and (MA)$_2$AgBiX$_6$ are quite similar due to their similar electronic configurations. Furthermore, because (MA)$_2$TlBiX$_6$ and MAPbX$_3$ are isoelectronic, their band structures are also very similar, as illustrated in Figure 3 below for X = Br$^-$.

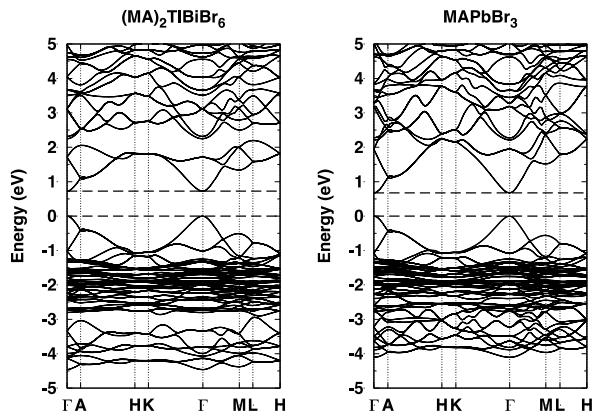

Figure 3 Computed electronic band structures of (MA)$_2$TlBiBr$_6$ and MAPbBr$_3$.

The mechanical stability of the hybrid double perovskites considered here is important if they are to be considered for real world device applications. Tables S6 and S7 give the calculated single crystal stiffness constants ($C_{ij}$) of the 15 structures and the corresponding polycrystalline values of Young's modulus (E), bulk modulus (B), shear modulus (G) and Poisson's ratio (ν). The stiffness constants are calculated from the stress-strain relationship by applying two types of strain to the unit cells,[28] $e_1$ and $e_3 + e_4$. $(MA)_2AgBiCl_6$, $(MA)_2AgBiBr_6$, $(MA)_2CuBiCl_6$ and $(MA)_2CuBiBr_6$ are found to be unstable when ±1% strain is applied due a rotation of the $MA^+$ cation making the stress-strain relationship nonlinear. Therefore, the stiffness constants are obtained using only ±0.5% strains, but even then the negative eigenvalues of the stiffness matrix indicate that $(MA)_2CuBiCl_6$ and $(MA)_2CuBiBr_6$ remain unstable.[29]

Figure 4 shows the polycrystalline elastic constants as a function of X anion radius, demonsrating that E, B and G decrease with increasing anion radius, in agreement with the trend observed from nano-indentation experiments on tetragonal $MAPbX_3$.[30] This can be explained by a decrease in the strength of the X⋯H and B-X bonds as the anion radius increases. However, it is found that the Young's moduli of the rhombohedral form of $MAPbX_3$ considered here are larger than the experimental results, probably because of the imposed symmetry constraints. The Poisson's ratio is largely independent of X anion radius. The directional dependencies of the single crystal Young's moduli of $(MA)_2KBiX_6$, $(MA)_2TlBiX_6$, $(MA)_2AgBiX_6$ and $MAPbX_3$ are shown in Figures S9-S12 and discussed in the SI.

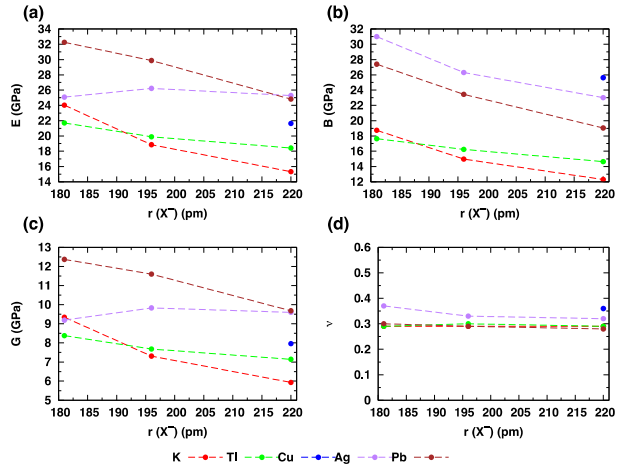

Figure 4 Polycrystalline values of (a) Young's modulus E, (b) bulk modulus B, (c) shear modulus G and (d) Poisson's ratio of $MA_2B^IB^{III}X_6$ as a function of the X anion radius.

A comparison between the Young's moduli of $MAPbX_3$ and $(MA)_2TlBiX_6$ shows that the isoelectronic substitution of Pb by Bi+Tl makes the structure less stiff, which is similar to findings in our previous studies on ZeoliticImidazolate Frameworks, when Zn is replaced by Li+B.[31] This can be explained by comparing the bond strength of Pb-X in $MAPbX_3$ with Bi-X and Tl-X in $(MA)_2TlBiX_6$. Assuming that bond strength and bond distance (d) are directly correlated, Figure S2 shows that $d_{Tl-X} > d_{Pb-X} > d_{Bi-X}$, which indicates that Bi-X is stronger than Pb-X and Tl-X. When applying strains, the weaker Tl-X bond results in a more flexible $(MA)_2TlBiX_6$ structure compared with $MAPbX_3$.

Motivated by the results from our DFT calculations, we have attempted to synthesize all the predicted phases using hydrothermal, solvent evaporation and solid state sintering methods. Despite the success with our previously reported $(MA)_2KBiCl_6$, we have only managed to obtain one other hybrid double perovskite. $(MA)_2TlBiBr_6$ was synthesized by the hydrothermal method at 150°C using 3mmolMABr, 1.5mmol $(CH_3COO)Tl$, and 1.5mmol $BiBr_3$ in 1ml HBr. The product was a mixture containing significant quantities of yellowish $(MA)_2Bi_3Br_9$. The double perovskite crystals are dark red in color, and can be easily separated manually from the mixture for further analysis. For the other compositions, although the appropriate starting stoichiometry was applied, the products were either $(MA)_3BiX_6$ (X = $Cl^-$) or $(MA)_3Bi_2X_9$ (X= $Br^-,I^-$). This could be because there is either an issue with kinetics during synthesis of the double perovskites or perhaps $(MA)_3Bi_2X_9$ is a thermodynamically favored phase.

$(MA)_2TlBiBr_6$, crystallizes in space group $Fm\overline{3}m$ (a = 11.762(2) Å), with $Tl^+$ and $Bi^{3+}$, isoelectronic with $Pb^{2+}$, occupy alternating octahedral sites. The $MA^+$ cations are disordered, and the corresponding electron density's size is ~2.25 Å, similar to the size of $MA^+$ (2.17 Å). Based on the size and the relatively isotropic electron density core, a rotational rather than translational motion of the organic cations in the cavity is suggested. A model wherein $MA^+$ is observed as an octahedron in the cavity with a restrained C-N length of 1.4Å was applied, as shown in Figure S13. A similar approach was used by Weller et. al. in their variable temperature neutron studies on $MAPbI_3$.[32] In this model, C and N occupy symmetry equivalent sites, each with a partial occupancy of 1/6. Structural refinement without $MA^+$ gives $R_{obs}$ = 6.31, with positive and negative residue electron densities of -2.65 and 2.29, respectively, while using this model $R_{obs}$ drops to 3.64 and the residues reduce to -2.17 and 1.85. A list of inorganic atomic coordinates is given in Table S8. The bond length of Bi-Br is 2.778(4) Å and that of Tl-Br is 3.103(4) Å, which are comparable with the calculated bond lengths of 2.854 Å and 3.145 Å, respectively, from our DFT calculations. The Pb-Br bond length is 2.965 Å in the previously reported cubic $MAPbBr_3$.[5]

Differential scanning calorimetry (DSC) was conducted on $(MA)_2TlBiBr_6$ using a DSC instrument Q2000 from 25°C to -150°C under liquid nitrogen flow of 100ml/min, with a ramping rate of 10°C/min. Two possible phase transitions are present, at approximately -10° C and -25°C (Figure S14). Based on our previous report on the $(MA)_2KBiCl_6$ analogue,[19] it can reasonably be predict that one of the low temperature phases would have rhombohedral symmetry. Low temperature single crystal X-ray diffraction, even at -20 °C, indicate heavy twinning, confirming that a phase transition takes place but making it very difficult to determine the structure. Further studies on the phase transitions will be carried out in the future using resonant ultrasound spectroscopy and variable temperature powder X-ray diffraction.

Optical measurements were carried out on $(MA)_2TlBiBr_6$ using a PerkinElmer Lambda 750 UV-Visible spectrometer in the reflectance mode with a 2nm slit width. The scan interval was 1 nm and the scan range was between 350 and 1000nm. By assuming a direct bandgap as suggested in the DFT calculation, a Tauc plot is constructed which estimates the optical band gap to be ~ 2.16 eV (Figure 5). This is comparable with that of $MAPbBr_3$ (2.2eV-2.35eV)[5,33] and is further consistent with the DFT calculations where the band structures of $(MA)_2TlBiX_6$ are predicted to be similar to the corresponding Pb analogues. Based on the structural and electronic analysis, the $(MA)_2TlBiX_6$ systems provide interesting alternatives to lead-containing perovskites, while noting that thalium itself is also toxic. The DFT calculations, which include SOC, underestimate the band gap for $(MA)_2TlBiBr_6$ (0.72 eV), as was the case for the previously reported $MAPbI_3$, with 0.60 eV from DFT with SOC compared to 1.55 eV from experiments.[34]

Nanoindentation experiments were performed on $(MA)_2TlBiX_6$ single crystals at room temperature, following the methods described previously.[30] A Young's modulus of 12.8 ± 1.9 GPa and hardness of 0.56 ± 0.10 GPa were observed with the indenter tip normal to the (111) facets in a cubic structure (see Figure S15). The experimental Young's modulus is consistent with the computational result (11.98 GPa) along the same direction (See Table S7).

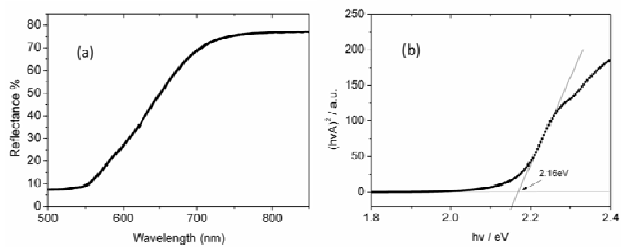

Figure 5 (a) Reflectance spectrum for $(MA)_2TlBiBr_6$, and (b) corresponding Tauc plot assuming a direct band gap.

## Conclusions

In conclusion, we have used DFT calculations to predict that some of the double perovskites of general formula $(MA)_2B^IBiX_6$ offer interesting alternatives to their lead-containing analogues, $MAPbX_3$, for photovoltaic applications, with band gaps and electronic structures that are similar to the lead-containing phases. Furthermore, we have synthesized a new double perovskite, $(MA)_2TlBiBr_6$, that is isoelectronic with $MAPbBr_3$ and has strikingly similar properties, Further reactions are being undertaken with the aim of obtaining additional phases in the hybrid double perovskite family.

## Acknowledgement

Z. Deng and S. Sun would like to thank the Cambridge Overseas Trust and the China Scholarship Council. F. Wei is a holder of an A*STAR international fellowship granted by the Agency for Science, Technology and Research,

Singapore. The calculations were performed at the Cambridge HPCS and the UK National Supercomputing Service, ARCHER. Access to the latter was obtained via the UKCP consortium and funded by EPSRC under Grant No. EP/K014560/1. G. Kieslich and A. K. Cheetham gratefully thank the Ras Al KaihmahCenter for Advanced Materials for financial support.

## Notes and references

1 M. D. McGehee, *Nat. Mater.*, 2014, **13**, 845–846.
2 A. Kojima, K. Teshima, Y. Shirai and T. Miyasaka, *J. Am. Chem. Soc.*, 2009, **131**, 6050–6051.
3 M. M. Lee, J. Teuscher, T. Miyasaka, T. N. Murakami and H. J. Snaith, *Science*, 2012, **338**, 643–647.
4 J. Burschka, N. Pellet, S.-J. Moon, R. Humphry-Baker, P. Gao, M. K. Nazeeruddin and M. Grätzel, *Nature*, 2013, **499**, 316–319.
5 T. Baikie, N. S. Barrow, Y. Fang, P. J. Keenan, P. R. Slater, R. O. Piltz, M. Gutmann, S. G. Mhaisalkar and T. J. White, *J. Mater. Chem. A*, 2015, **3**, 9298–9307.
6 A. Jaffe, Y. Lin, C. M. Beavers, J. Voss, W. L. Mao and H. I. Karunadasa, *ACS Cent. Sci.*, 2016, **2**, 201–209.
7 J. M. Frost and A. Walsh, *Acc. Chem. Res.*, 2016, **49**, 528–535.
8 A. L. Montero-Alejo, E. Menéndez-Proupin, D. Hidalgo-Rojas, P. Palacios, P. Wahnón and J. C. Conesa, *J. Phys. Chem. C*, 2016, **120**, 7976–7986.
9 T. M. Brenner, D. A. Egger, A. M. Rappe, L. Kronik, G. Hodes and D. Cahen, *J. Phys. Chem. Lett.*, 2015, **6**, 4754–4757.
10 L. M. Pazos-Outon, M. Szumilo, R. Lamboll, J. M. Richter, M. Crespo-Quesada, M. Abdi-Jalebi, H. J. Beeson, M. Vruini, M. Alsari, H. J. Snaith, B. Ehrler, R. H. Friend and F. Deschler, *Science*, 2016, **351**, 1430–1433.
11 S. Yang, Y. Wang, P. Liu, Y.-B. Cheng, H. J. Zhao and H. G. Yang, *Nat. Energy*, 2016, **1**, 15016.
12 A. J. Lehner, D. H. Fabini, H. A. Evans, C.-A. Hébert, S. R. Smock, J. Hu, H. Wang, J. W. Zwanziger, M. L. Chabinyc and R. Seshadri, *Chem. Mater.*, 2015, **27**, 7137–7148.
13 B. Chabot and E. Parthé, *Acta Crystallogr. Sect. B Struct. Crystallogr. Cryst. Chem.*, 1978, **34**, 645–648.
14 S. Sun, S. Tominaka, J.-H. Lee, F. Xie, P. D. Bristowe and A. K. Cheetham, *APL Mater.*, 2016, **4**, 031101.
15 A. H. Slavney, T. Hu, A. M. Lindenberg and H. I. Karunadasa, *J. Am. Chem. Soc.*, 2016, **138**, 2138–2141.
16 G. Volonakis, M. R. Filip, A. A. Haghighirad, N. Sakai, B. Wenger, H. J. Snaith and F. Giustino, *J. Phys. Chem. Lett.*, 2016, **7**, 1254–1259.
17 E. T. McClure, M. R. Ball, W. Windl and P. M. Woodward, *Chem. Mater.*, 2016, **28**, 1348–1354.
18 L. R. Morss, M. Siegal, L. Stenger and N. Edelstein, *Inorg. Chem.*, 1970, **9**, 1771–1775.
19 F. Wei, Z. Deng, S. Sun, F. Xie, G. Kieslich, D. M. Evans, M. A. Carpenter, P. D. Bristowe and A. K. Cheetham, *Mater. Horiz.*, 2016, DOI: 10.1039/c6mh00053c.
20 K. Momma and F. Izumi, *J. Appl. Crystallogr.*, 2011, **44**, 1272–1276.
21 G. Kresse, *J. Non. Cryst. Solids*, 1995, **192-193**, 222–229.
22 G. Kresse and J. Furthmüller, *Comput. Mater. Sci.*, 1996, **6**, 15–50.
23 R. D. Shannon, *Acta Crystallogr. Sect. A*, 1976, **32**, 751–767.
24 J.-H. Lee, N. C. Bristowe, P. D. Bristowe and A. K. Cheetham, *Chem. Commun.*, 2015, **51**, 6434–6437.
25 J. Even, L. Pedesseau, J.-M. Jancu and C. Katan, *J. Phys. Chem. Lett.*, 2013, **4**, 2999–3005.
26 J.-H. Lee, N. C. Bristowe, J. H. Lee, S.-H. Lee, P. D. Bristowe, A. K. Cheetham and H. M. Jang, *Chem. Mater.*, 2016, DOI: acs.chemmater.6b00968.
27 A. Filippetti and A. Mattoni, *Phys. Rev. B*, 2014, **89**, 125203.
28 Y. Le Page and P. Saxe, *Phys. Rev. B*, 2002, **65**, 104104.
29 F. Mouhat and F.-X. Coudert, *Phys. Rev. B*, 2014, **90**, 224104.
30 S. Sun, Y. Fang, G. Kieslich, T. J. White and A. K. Cheetham, *J. Mater. Chem. A*, 2015, **3**, 18450–18455.
31 T. D. Bennett, J.-C. Tan, S. A. Moggach, R. Galvelis, C. Mellot-Draznieks, B. A. Reisner, A. Thirumurugan, D. R. Allan and A. K. Cheetham, *Chem. - A Eur. J.*, 2010, **16**, 10684–10690.
32 M. T. Weller, O. J. Weber, P. F. Henry, M. Di Pumpo and T. C. Hansen, *Chem. Commun.*, 2015, **51**, 3–6.
33 H. Bin Kim, I. Im, Y. Yoon, S. Do Sung, E. Kim, J. Kim and W. I. Lee, *J. Mater. Chem. A*, 2015, **3**, 9264–9270.
34 P. Umari, E. Mosconi and F. De Angelis, *Sci. Rep.*, 2014, **4**, 4467.

# Exploring the properties of lead-free hybrid double perovskites using a combined computational-experimental approach

Zeyu Deng, Fengxia Wei, Shijing Sun, GregorKieslich, Anthony K. Cheetham* and Paul D. Bristowe*

Department of Materials Science and Metallurgy

University of Cambridge

27 Charles Babbage Road

Cambridge CB3 0FS

United Kingdom

**Computational Methodology**

The DFT calculations were performed using the Vienna *ab initio* Simulation Package (VASP).[1,2] Projected augmented wave (PAW)[3] pseudopotentials were employed with the following electrons treated explicitly: H ($1s^1$), C ($2s^22p^2$), N ($2s^22p^3$), K ($3s^23p^64s^1$), Bi ($5d^{10}6s^26p^3$), Tl ($5d^{10}6s^26p^1$), Cu ($3p^63d^{10}4s^1$), Ag ($4p^64d^{10}5s^1$), Pb ($5d^{10}6s^26p^2$), Cl ($3s^23p^5$), Br ($4s^24p^5$) and I ($5s^25p^5$). The non-local van der Waals density functional (vdW-DF)[4] was used with the exchange-correlation energy calculated as $Exc = E_x^{GGA} + E_c^{LDA} + E_c^{nl}$, where the exchange energy $E_x^{GGA}$ is obtained from the generalized gradient approximation (GGA) using the optB86b functional, the local correlation energy $E_c^{LDA}$ from the local density approximation (LDA) and the non-local correlation energy $E_c^{nl}$ from double space integration. K-points were sampled in the first Brillouin zone using a 4×4×2 Monkhorst-Pack[5] mesh, and for electronic density of states (DOS) calculations, a finer 8×8×3 mesh was used. A 500 eV plane-wave kinetic energy cutoff was employed for all calculations. The effect of relativistic spin-orbit coupling (SOC) was included in the DOS and electronic band structure calculations. The experimentally synthesized double perovskite $(MA)_2KBiCl_6$ was used as a basis for constructing all the structures which were then relaxed until the interatomic forces were less than 0.01 eV/Å while maintaining the same rhombohedral symmetry as $(MA)_2KBiCl_6$.

1   G. Kresse, *J. Non. Cryst. Solids*, 1995, **192-193**, 222–229.
2   G. Kresse and J. Furthmüller, *Comput. Mater. Sci.*, 1996, **6**, 15–50.
3   P. E. Blochl, *Phys Rev B Condens Matter*, 1994, **50**, 17953-17979.
4   J. Klimes, D. R. Bowler and A. Michaelides, *Phys Rev B*, 2011, **83**.
5   H. J. Monkhorst and J. D. Pack, *Phys Rev B*, 1976, **13**, 5188-5192.

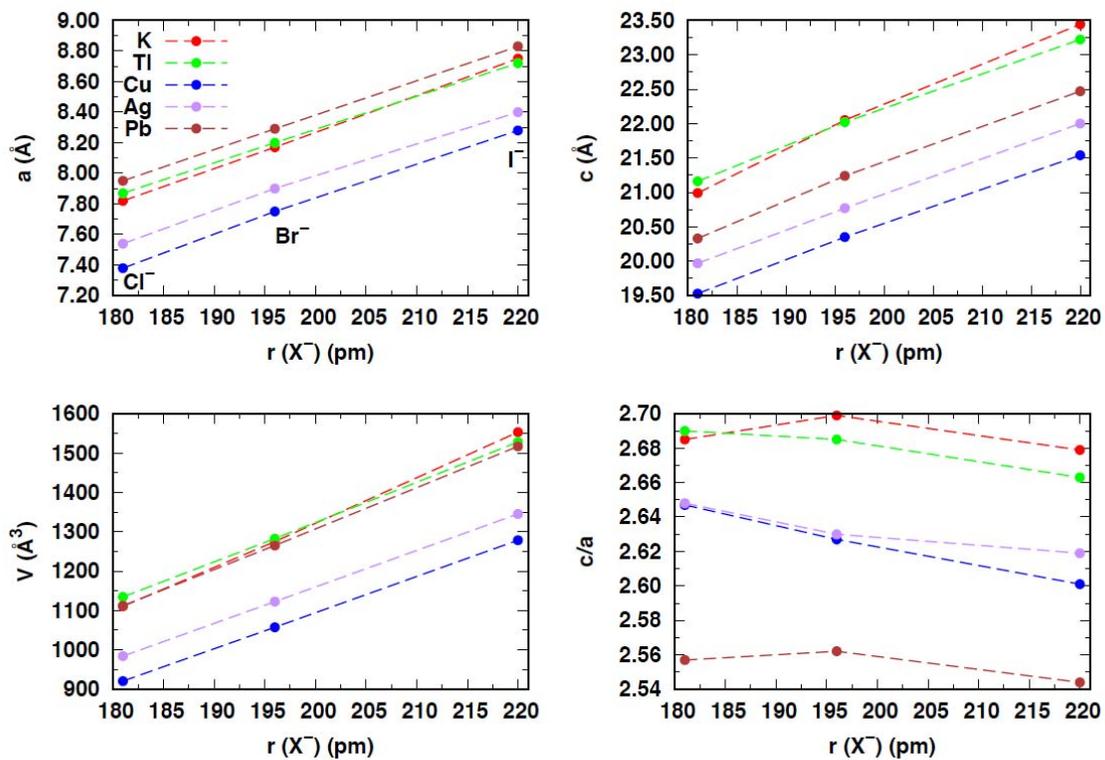

**Figure S1** Computed lattice constants, equilibrium volumes and c/a ratios of $(MA)_2B^IB^{III}X_6$ as a function of the radius of anion X.

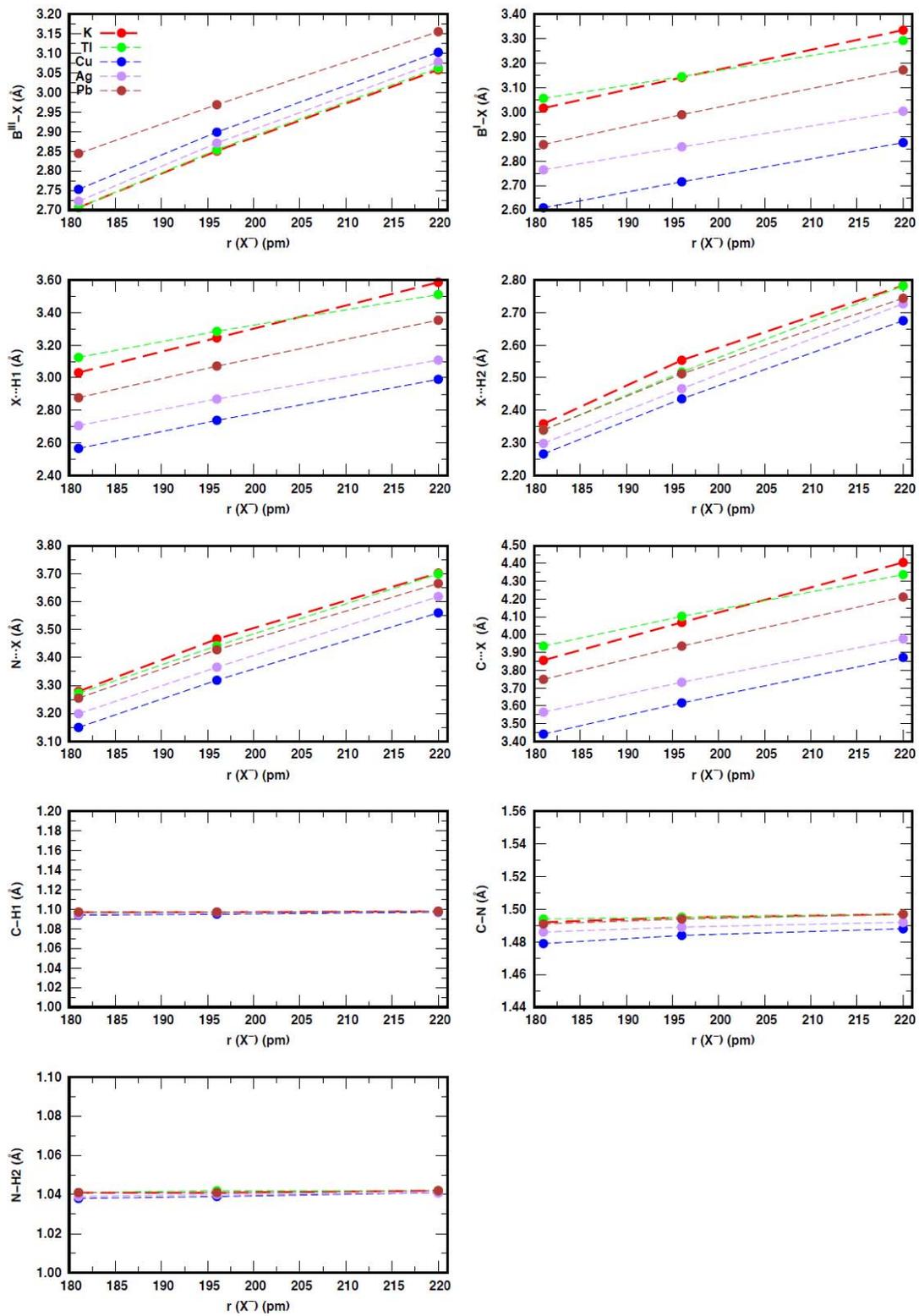

**Figure S2** Computed interatomic distances in $(MA)_2B^IB^{III}X_6$ as a function of the radius of anion X.

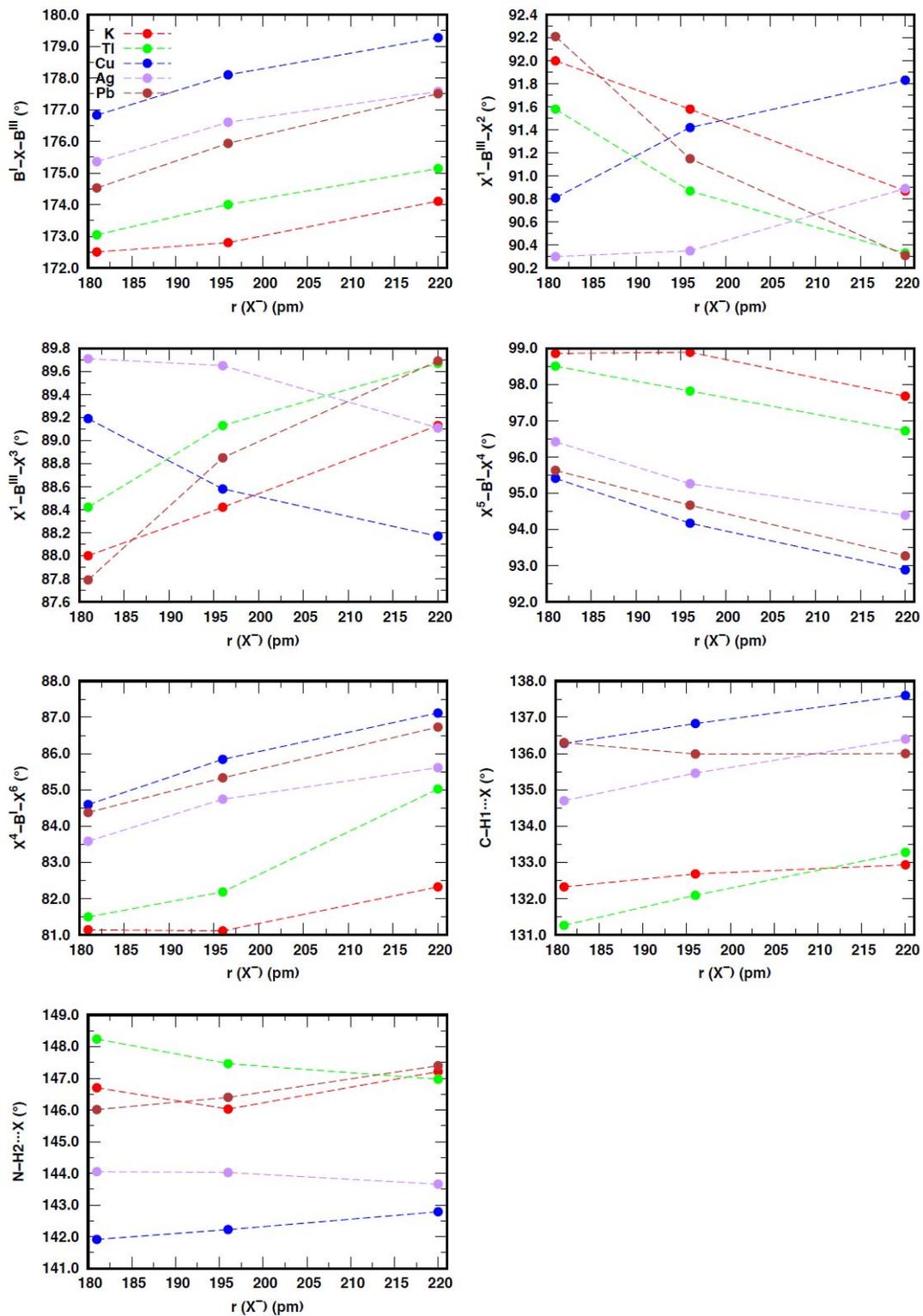

**Figure S3** Computed bond angles in $(MA)_2B^IB^{III}X_6$ as a function of the radius of anion X.

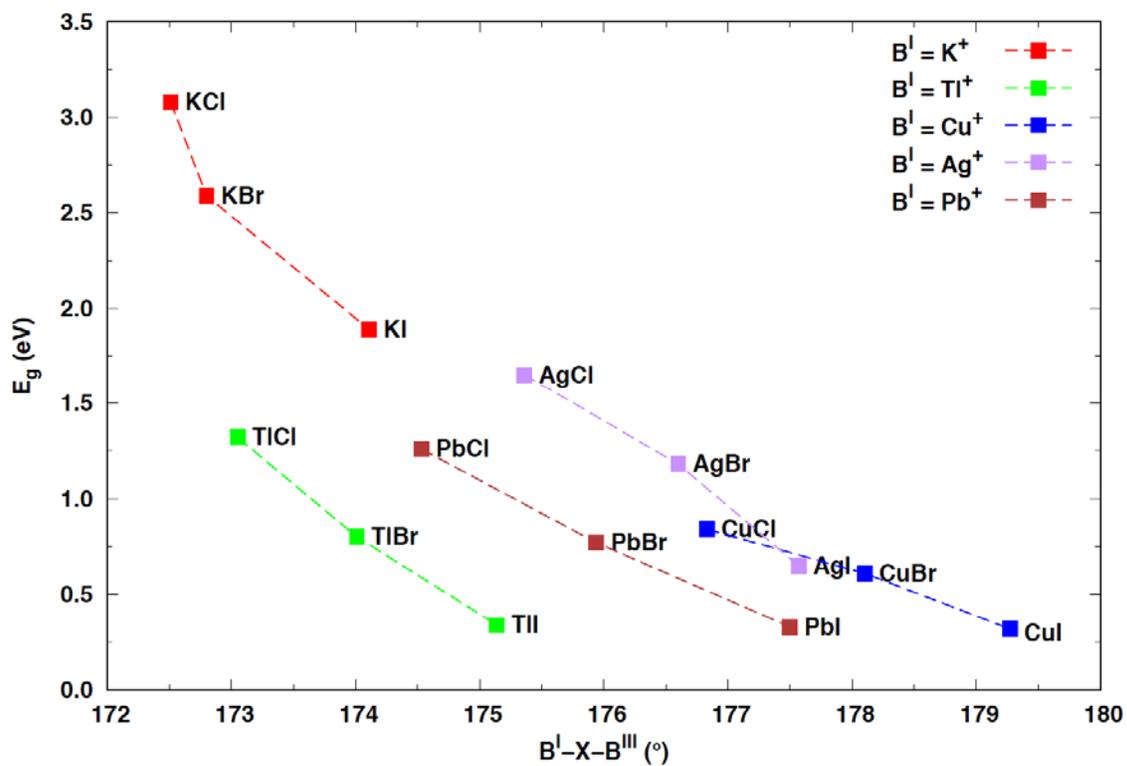

**Figure S4** Computed band gaps of $(MA)_2B^IB^{III}X_6$ as a function of the $B^I$-X-$B^{III}$ bond angle.

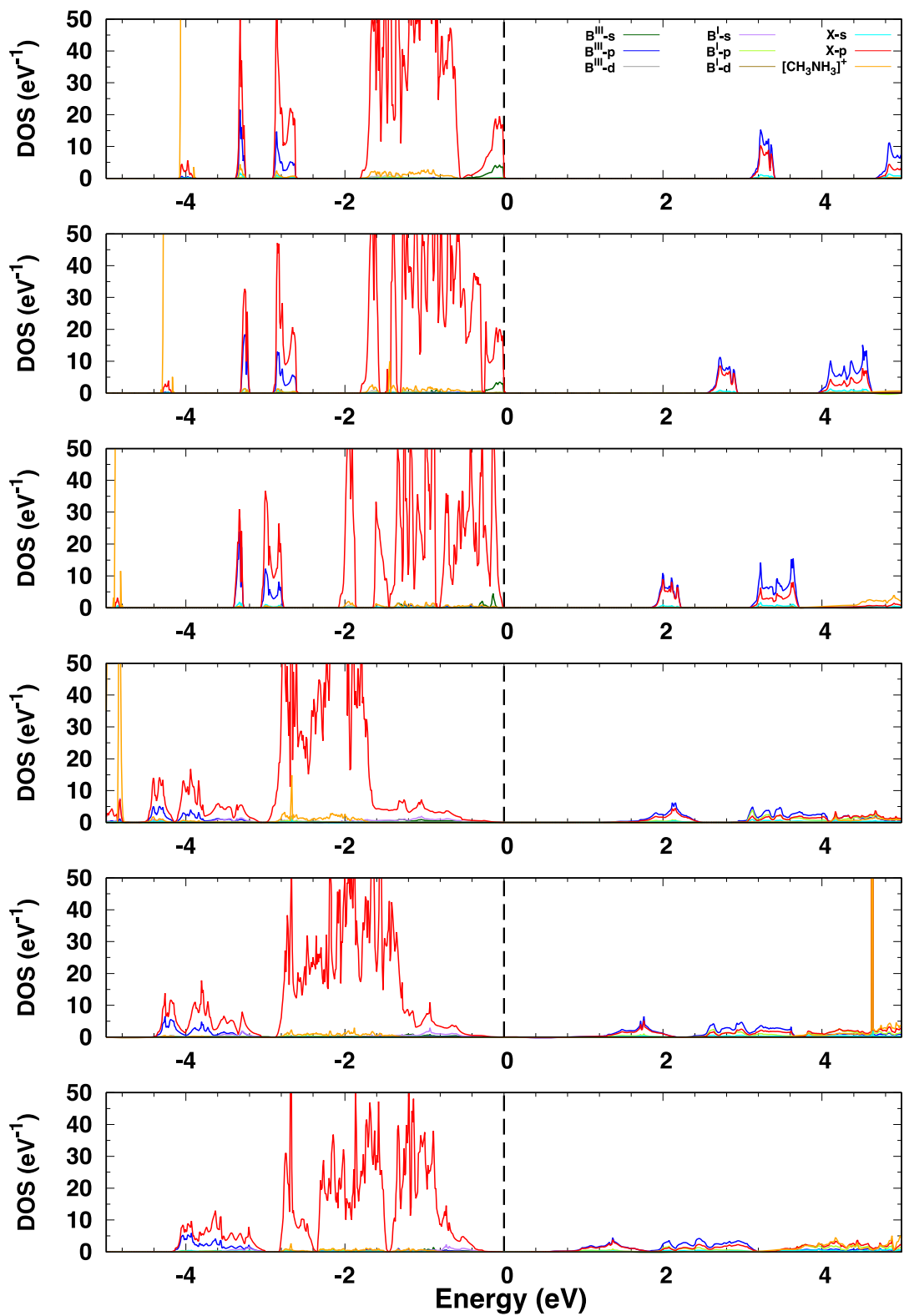

**Figure S5** Computed PDOS of $(MA)_2B^IB^{III}X_6$, where $B^{III}$ = Bi, $B^I$ = K and Tl, X = Cl, Br and I. From top to bottom: $(MA)_2KBiCl_6$, $(MA)_2KBiBr_6$, $(MA)_2KBiI_6$, $(MA)_2TlBiCl_6$, $(MA)_2TlBiBr_6$ and $(MA)_2TlBiI_6$.

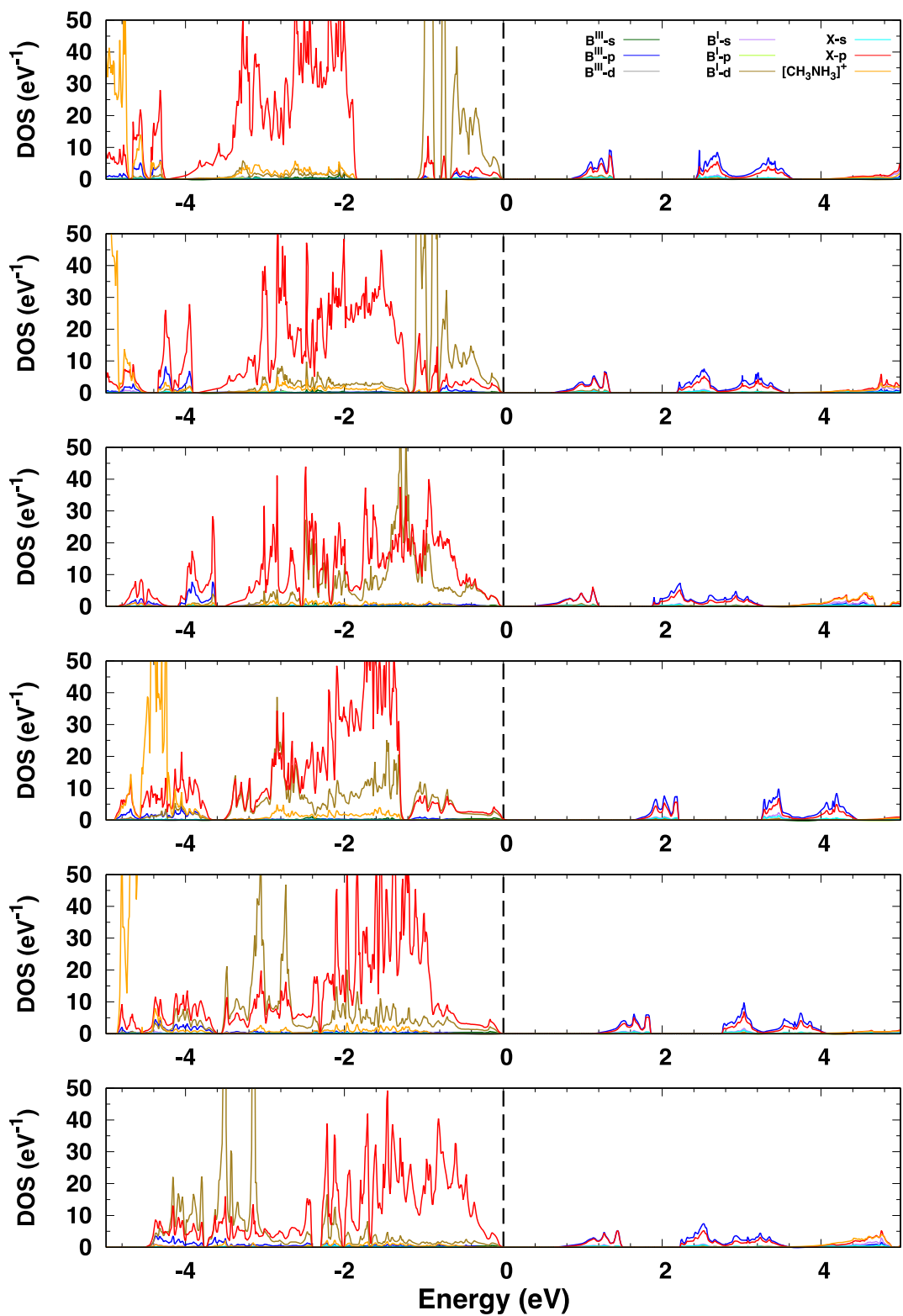

**Figure S6** Computed PDOS of (MA)$_2$B$^I$B$^{III}$X$_6$, where B$^{III}$ = Bi, B$^I$ = Cu and Ag, X = Cl, Br and I. From top to bottom: (MA)$_2$CuBiCl$_6$, (MA)$_2$CuBiBr$_6$, (MA)$_2$CuBiI$_6$, (MA)$_2$AgBiCl$_6$, (MA)$_2$AgBiBr$_6$ and (MA)$_2$AgBiI$_6$.

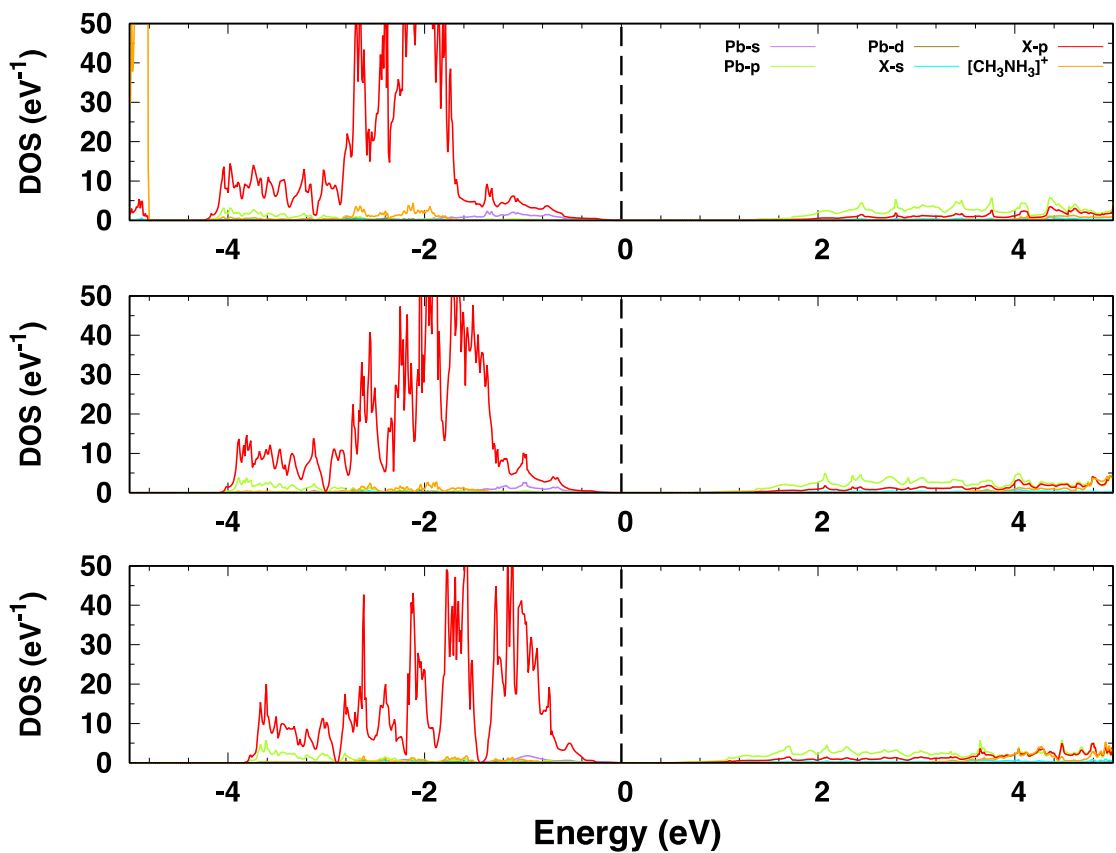

**Figure S7** Computed PDOS of MAPbX$_3$. From top to bottom: MAPbCl$_3$, MAPbBr$_3$ and MAPbI$_3$.

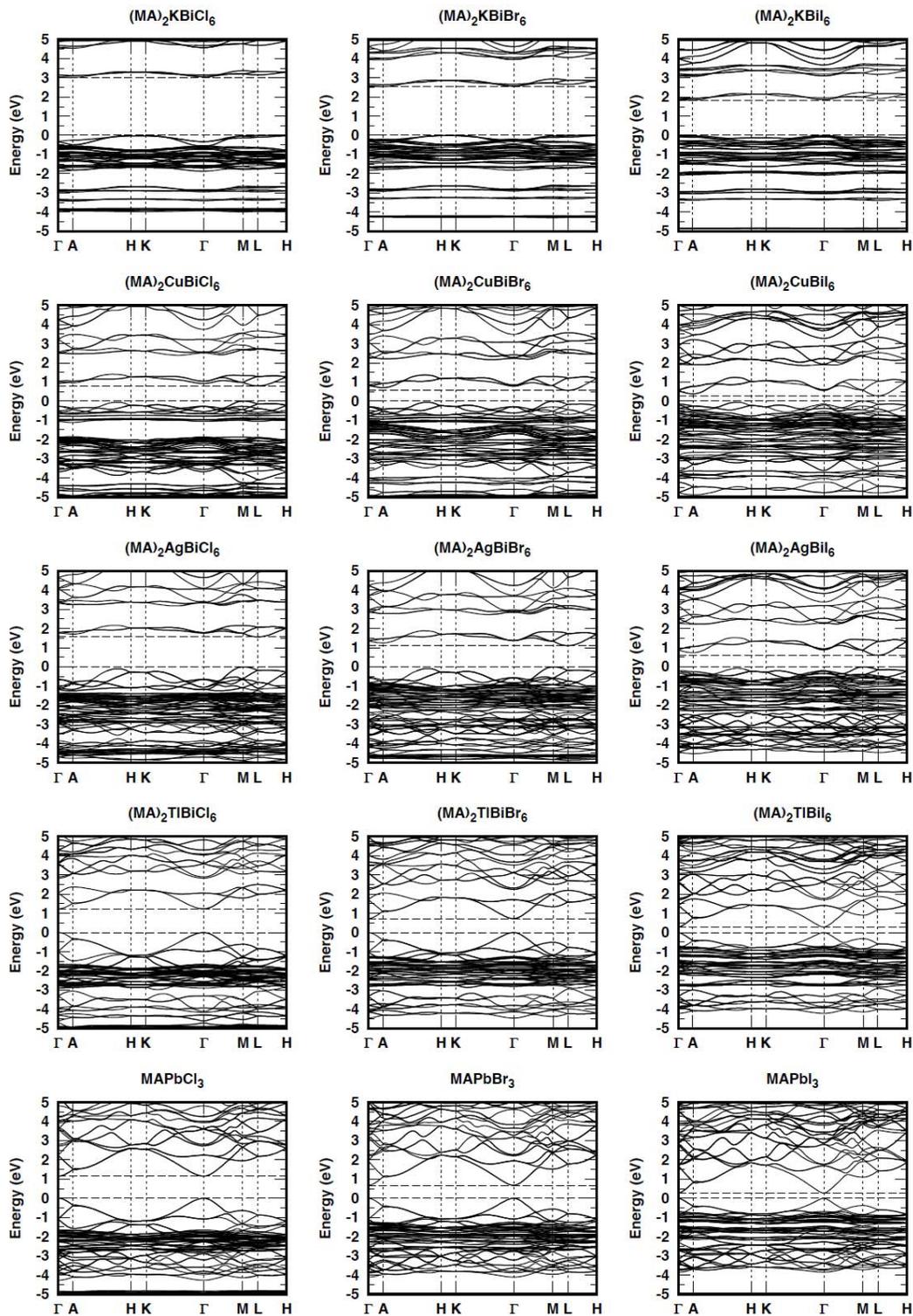

**Figure S8** Computed band structures of $(MA)_2B^IB^{III}X_6$ including $MAPbX_3$. The following high symmetry points in the first Brillouin zone were used: Γ (0,0,0), A (0,0,0.5), H (-0.333,0.667,0.5), K (-0.333,0.667,0), M (0,0.5,0) and L (0,0.5,0.5).

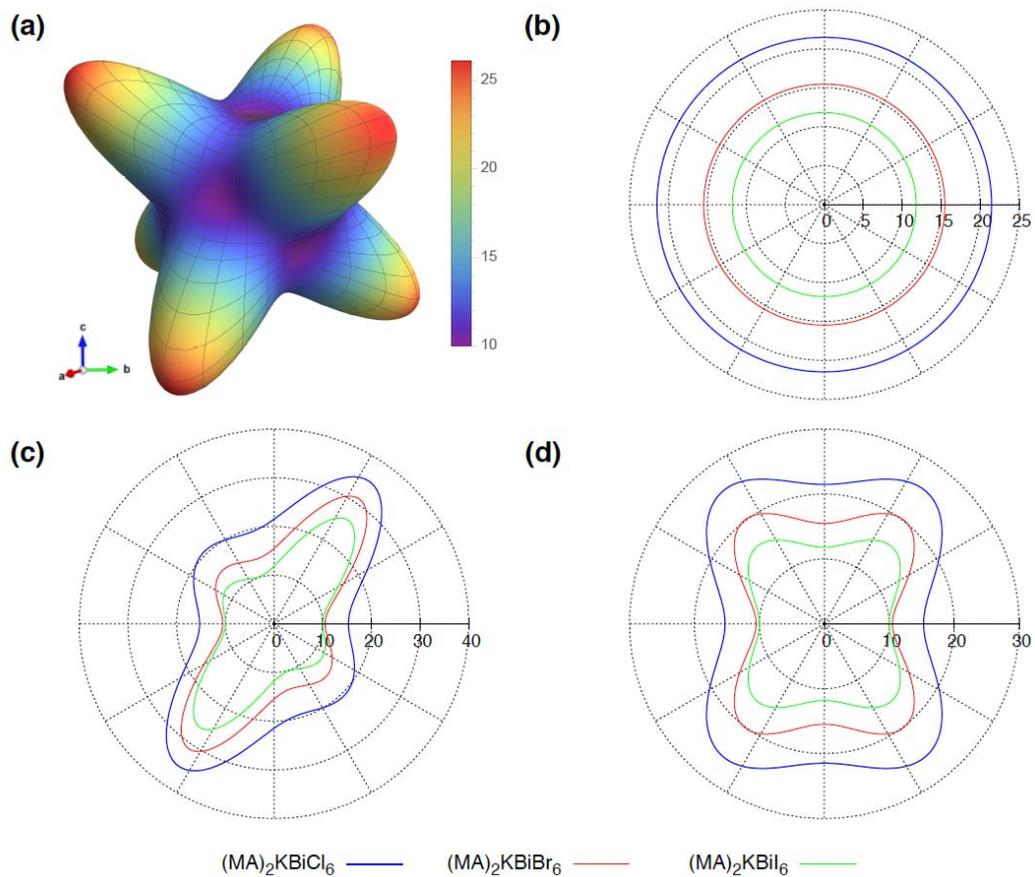

**Figure S9** (a) Computed 3D directional Young's modulus of $(MA)_2KBiI_6$ and the corresponding contour plots for $(MA)_2KBiX_6$ on (b) the (001) plane (c) the plane perpendicular to [100] and (d) the (010) plane. Units shown are in GPa.

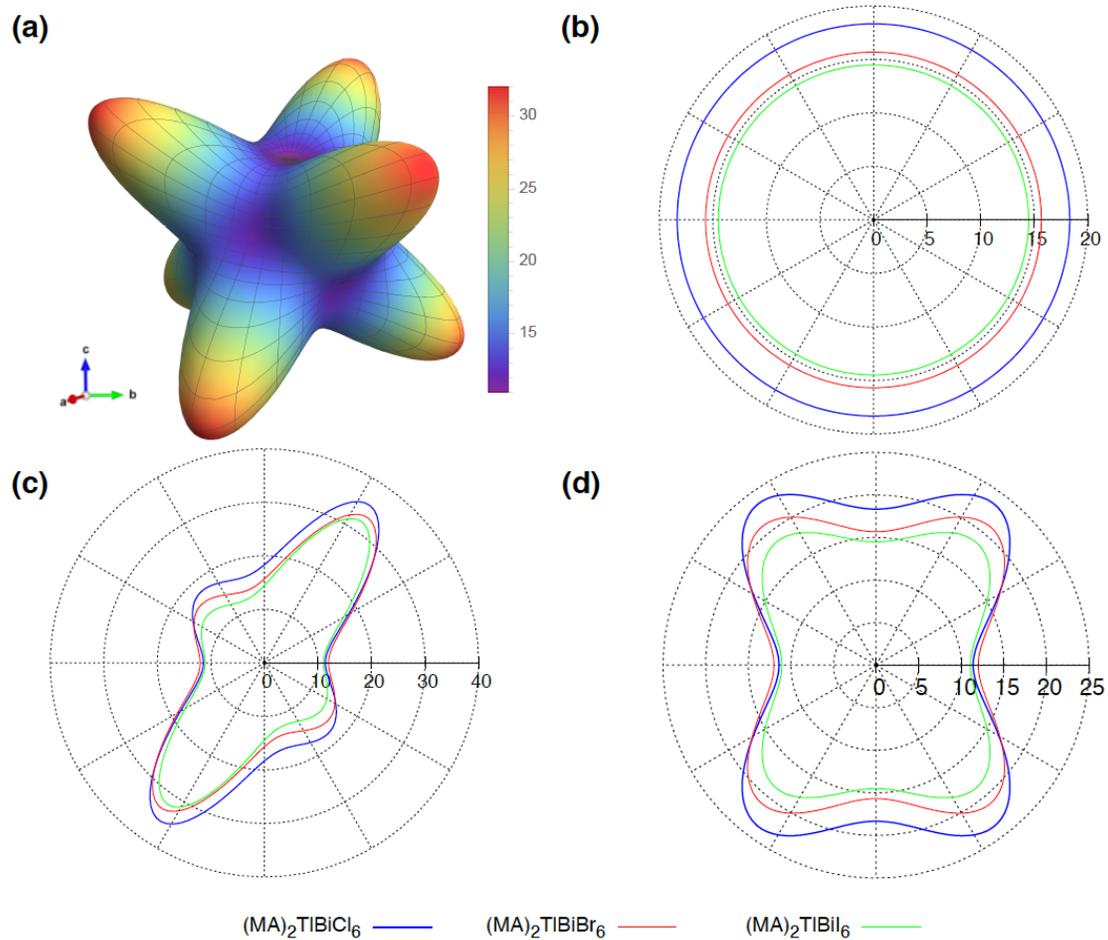

**Figure S10**(a) Computed 3D directional Young's modulus of $(MA)_2TlBiI_6$ and the corresponding contour plots for $(MA)_2TlBiX_6$ on (b) the (001) plane (c) the plane perpendicular to [100] and (d) the (010) plane. Units shown are in GPa.

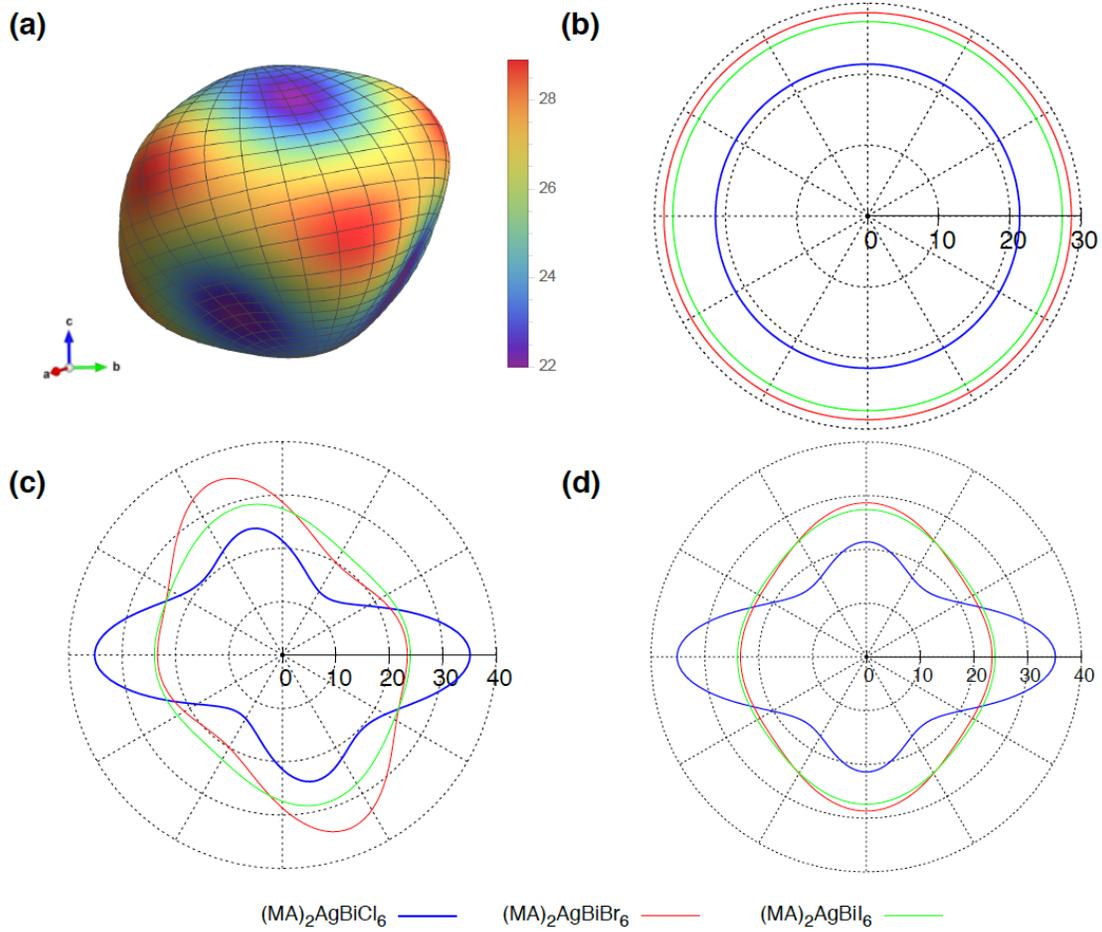

**Figure S11**(a) Computed 3D directional Young's modulus of $(MA)_2AgBiI_6$ and the corresponding contour plots for $(MA)_2AgBiX_6$ on (b) the (001) plane (c) the plane perpendicular to [100] and (d) the (010) plane. Units shown are in GPa.

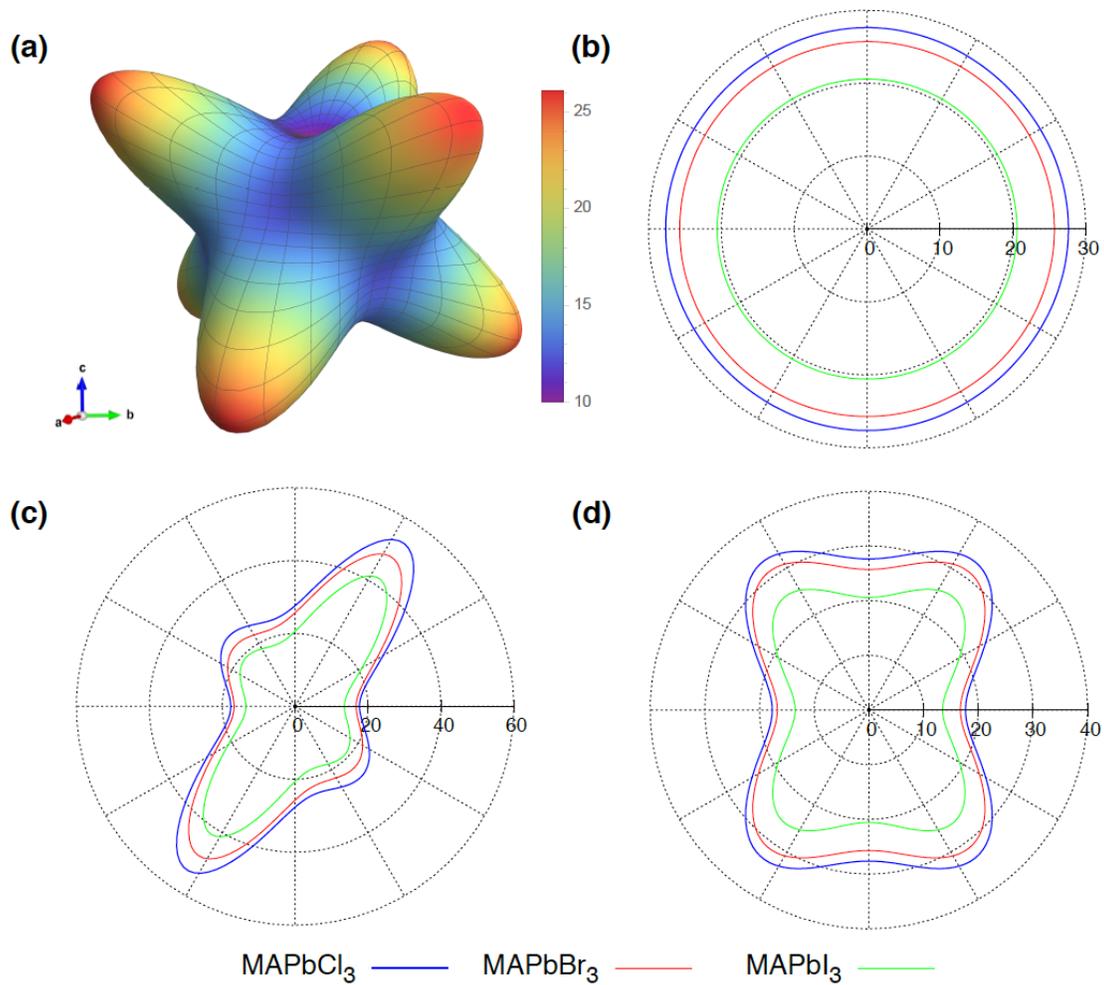

**Figure S12**(a) Computed 3D directional Young's modulus of MAPbI$_3$ and the corresponding contour plotsfor MAPbX$_3$on (b) the (001) plane (c) the plane perpendicular to [100]and (d)the (010) plane. Units shown are in GPa.

**Notes on the computed Young's moduli**

For $(MA)_2KBiX_6$, $(MA)_2TlBiX_6$ and $MAPbX_3$, the maximum and minimum values lie along B-X bond directions and the diagonal of the B-X cage, respectively. However, for $(MA)_2AgBiX_6$ (and $(MA)_2CuBiI_6$ not shown) these maxima and minima reverse directions. This can be explained in terms of the incompressibility of the $MA^+$ cation and the Goldschmidt tolerance factors (see Table S8).[6] Previous experimental studies on $MAPbCl_3$ showed that the most rigid component of the structure is the $MA^+$ cation rather than the $PbCl_6$ octahedra.[7] Also, as Figure S2 shows, when incorporating different X and $B^I$ ions, most bond lengths change by around 0.3 ~ 0.4 Å whereas the change in C-N bond length is much smaller at less than 0.01 Å. The tolerance factor for $(MA)_2AgBiX_6$ is relatively large, implying that the $B^IXB^{III}$ framework is too small for the $MA^+$ cation. Therefore, because $MA^+$ is incompressible and the framework is too small, when applying strains, the intramolecular interactions between $MA^+$ and the framework either make $MA^+$ rotate, or compress the C-N, C-H or N-H bonds, inducing a large change in stress. The $MA^+$ rotations result in the nonlinear stress-strain relationship found in the calculations for those structures with large tolerance factors, and the bond compressions make the Young's moduli along the C-N, C-H or N-H directions exhibit a maximum.

6  Kieslich, G.; Sun, S.; Cheetham, A. K. *Chem. Sci.* **2014**, *5*, 4712–4715.

7   Chi, L.; Swainson, I.; Cranswick, L.; Her, J.-H.; Stephens, P.; Knop, O. *J. Solid State Chem.* **2005**, *178*, 1376–1385.

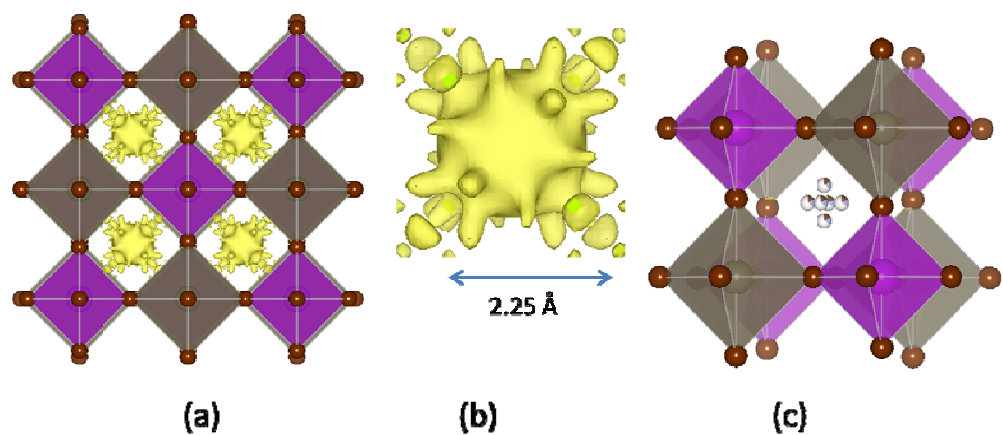

**Figure S13** (a) Crystal structure of (MA)$_2$TlBiBr$_6$ obtained from single crystal X-ray diffraction. TlBr$_6$ and BiBr$_6$ octahedra are grey and purple respectively. MA$^+$ cation is shown as the electron density (b) the shape of the electron density and (c) MA$^+$ cation is modeled as the octahedron inside the cavity with restrained C-N bond length 1.4 Å.

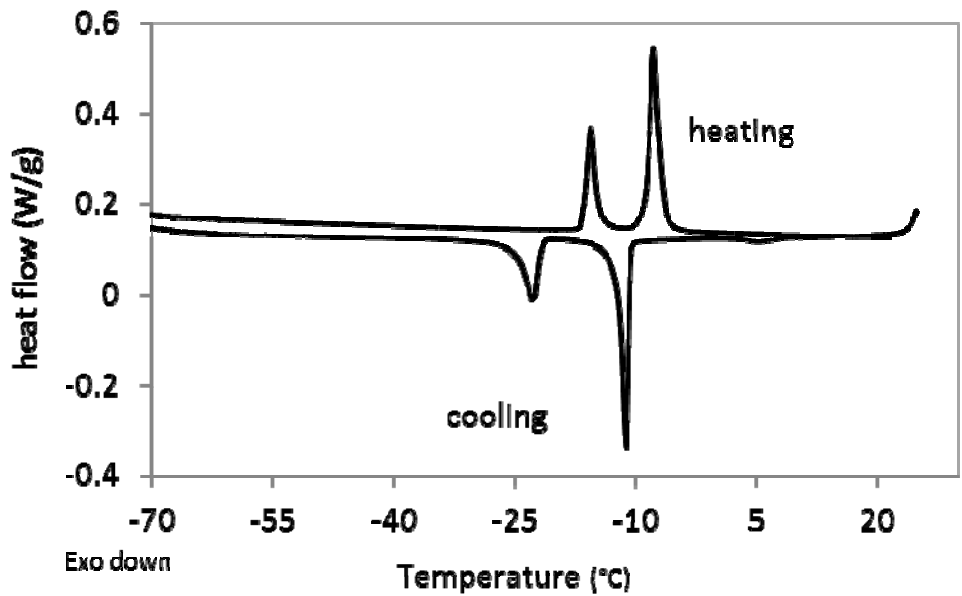

**Figure S14** DSC curve of (MA)$_2$TlBiBr$_6$.

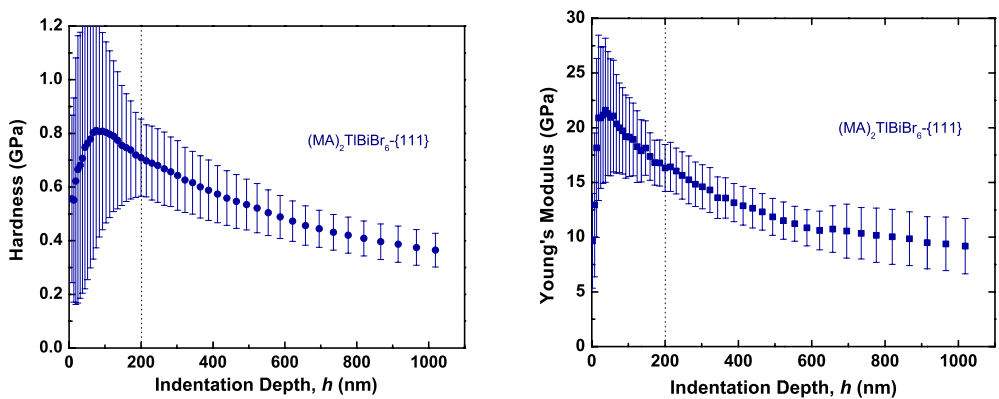

**Figure S15** Hardness and Young's Modulus for (MA)$_2$TlBiBr$_6$ from nanoindentation.

**Table S1** Computed lattice constants (Å), equilibrium volumes (Å$^3$) and c/a ratios of (MA)$_2$B$^I$B$^{III}$X$_6$

| A | B$^I$ | B$^{III}$ | X | a | c | V | c/a |
|---|---|---|---|---|---|---|---|
| MA | K | Bi | Cl | 7.82 | 20.99 | 1110.64 | 2.69 |
| MA | K | Bi | Br | 8.17 | 22.05 | 1274.99 | 2.70 |
| MA | K | Bi | I | 8.75 | 23.44 | 1553.34 | 2.68 |
| MA | Tl | Bi | Cl | 7.87 | 21.16 | 1134.52 | 2.69 |
| MA | Tl | Bi | Br | 8.20 | 22.02 | 1282.42 | 2.69 |
| MA | Tl | Bi | I | 8.72 | 23.22 | 1528.19 | 2.66 |
| MA | Cu | Bi | Cl | 7.38 | 19.53 | 920.57 | 2.65 |
| MA | Cu | Bi | Br | 7.75 | 20.35 | 1057.67 | 2.63 |
| MA | Cu | Bi | I | 8.28 | 21.54 | 1278.68 | 2.60 |
| MA | Ag | Bi | Cl | 7.54 | 19.97 | 984.27 | 2.65 |
| MA | Ag | Bi | Br | 7.90 | 20.77 | 1122.54 | 2.63 |
| MA | Ag | Bi | I | 8.40 | 22.00 | 1345.23 | 2.62 |
| MA | Pb | Pb | Cl | 7.95 | 20.33 | 1112.42 | 2.56 |
| MA | Pb | Pb | Br | 8.29 | 21.24 | 1265.21 | 2.56 |
| MA | Pb | Pb | I | 8.83 | 22.47 | 1517.20 | 2.54 |

**Table S2** Computed interatomic distances (Å) of $(MA)_2B^IB^{III}X_6$

| A | $B^I$ | $B^{III}$ | X | $B^{III}$-X | $B^I$-X | X⋯H1 | X⋯H2 | N⋯X | C⋯X | C-N | C-H1 | N-H2 |
|---|---|---|---|---|---|---|---|---|---|---|---|---|
| MA | K | Bi | Cl | 2.706 | 3.016 | 3.032 | 2.359 | 3.279 | 3.856 | 1.492 | 1.097 | 1.041 |
| MA | K | Bi | Br | 2.851 | 3.141 | 3.245 | 2.554 | 3.466 | 4.069 | 1.495 | 1.097 | 1.041 |
| MA | K | Bi | I | 3.058 | 3.335 | 3.585 | 2.783 | 3.702 | 4.406 | 1.497 | 1.098 | 1.042 |
| MA | Tl | Bi | Cl | 2.706 | 3.056 | 3.125 | 2.340 | 3.272 | 3.936 | 1.494 | 1.097 | 1.041 |
| MA | Tl | Bi | Br | 2.854 | 3.145 | 3.286 | 2.518 | 3.442 | 4.103 | 1.495 | 1.097 | 1.042 |
| MA | Tl | Bi | I | 3.063 | 3.292 | 3.511 | 2.782 | 3.699 | 4.337 | 1.497 | 1.098 | 1.042 |
| MA | Cu | Bi | Cl | 2.753 | 2.609 | 2.567 | 2.267 | 3.150 | 3.441 | 1.479 | 1.094 | 1.038 |
| MA | Cu | Bi | Br | 2.899 | 2.715 | 2.739 | 2.436 | 3.319 | 3.617 | 1.484 | 1.095 | 1.039 |
| MA | Cu | Bi | I | 3.103 | 2.875 | 2.990 | 2.675 | 3.560 | 3.872 | 1.488 | 1.097 | 1.041 |
| MA | Ag | Bi | Cl | 2.722 | 2.764 | 2.708 | 2.299 | 3.199 | 3.565 | 1.486 | 1.095 | 1.039 |
| MA | Ag | Bi | Br | 2.871 | 2.858 | 2.871 | 2.467 | 3.365 | 3.733 | 1.489 | 1.097 | 1.040 |
| MA | Ag | Bi | I | 3.078 | 3.004 | 3.109 | 2.727 | 3.618 | 3.976 | 1.492 | 1.098 | 1.041 |
| MA | Pb | Pb | Cl | 2.844 | 2.868 | 2.878 | 2.340 | 3.255 | 3.748 | 1.491 | 1.097 | 1.041 |
| MA | Pb | Pb | Br | 2.969 | 2.990 | 3.072 | 2.512 | 3.428 | 3.936 | 1.494 | 1.097 | 1.041 |
| MA | Pb | Pb | I | 3.155 | 3.172 | 3.353 | 2.744 | 3.665 | 4.212 | 1.497 | 1.098 | 1.042 |

**Table S3** Computed bond angles (°) in $(MA)_2B^IB^{III}X_6$.

| A | $B^I$ | $B^{III}$ | X | $B^I$-X-$B^{III}$ | $X^1$-$B^{III}$-$X^2$ | $X^1$-$B^{III}$-$X^2$ | $X^5$-$B^I$-$X^4$ | $X^4$-$B^I$-$X^6$ | C-H1···X | N-H2···X |
|---|---|---|---|---|---|---|---|---|---|---|
| MA | K | Bi | Cl | 172.51 | 92.00 | 88.00 | 98.86 | 81.14 | 132.32 | 146.70 |
| MA | K | Bi | Br | 172.80 | 91.58 | 88.42 | 98.89 | 81.11 | 132.68 | 146.03 |
| MA | K | Bi | I | 174.11 | 90.87 | 89.13 | 97.68 | 82.32 | 132.93 | 147.21 |
| MA | Tl | Bi | Cl | 173.05 | 91.58 | 88.42 | 98.51 | 81.49 | 131.26 | 148.24 |
| MA | Tl | Bi | Br | 174.01 | 90.87 | 89.13 | 97.82 | 82.18 | 132.09 | 147.46 |
| MA | Tl | Bi | I | 175.14 | 90.33 | 89.67 | 96.72 | 85.02 | 133.27 | 146.97 |
| MA | Cu | Bi | Cl | 176.83 | 90.81 | 89.19 | 95.41 | 84.59 | 136.28 | 141.92 |
| MA | Cu | Bi | Br | 178.10 | 91.42 | 88.58 | 94.17 | 85.84 | 136.83 | 142.23 |
| MA | Cu | Bi | I | 179.27 | 91.83 | 88.17 | 92.88 | 87.12 | 137.60 | 142.79 |
| MA | Ag | Bi | Cl | 175.36 | 90.30 | 89.71 | 96.42 | 83.58 | 134.70 | 144.05 |
| MA | Ag | Bi | Br | 176.60 | 90.35 | 89.65 | 95.26 | 84.74 | 135.46 | 144.03 |
| MA | Ag | Bi | I | 177.57 | 90.89 | 89.11 | 94.39 | 85.61 | 136.40 | 143.66 |
| MA | Pb | Pb | Cl | 174.53 | 92.21 | 87.79 | 95.63 | 84.37 | 136.30 | 146.01 |
| MA | Pb | Pb | Br | 175.94 | 91.15 | 88.85 | 94.67 | 85.33 | 135.99 | 146.40 |
| MA | Pb | Pb | I | 177.50 | 90.31 | 89.69 | 93.27 | 86.73 | 136.00 | 147.40 |

**Table S4** Computed band gaps (eV) of $(MA)_2B^IB^{III}X_6$ with experimental values for $(MA)_2KBiCl_6$ and $(MA)_2TlBiBr_6$.

| A | $B^I$ | $B^{III}$ | X | $E_g$ (DFT+SOC) | $E_g$ (exp) |
|---|---|---|---|---|---|
| MA | K | Bi | Cl | 3.02 | 3.04 |
| MA | K | Bi | Br | 2.54 | |
| MA | K | Bi | I | 1.84 | |
| MA | Tl | Bi | Cl | 1.23 | |
| MA | Tl | Bi | Br | 0.72 | 2.16 |
| MA | Tl | Bi | I | 0.28 | |
| MA | Cu | Bi | Cl | 0.79 | |
| MA | Cu | Bi | Br | 0.56 | |
| MA | Cu | Bi | I | 0.28 | |
| MA | Ag | Bi | Cl | 1.57 | |
| MA | Ag | Bi | Br | 1.11 | |
| MA | Ag | Bi | I | 0.60 | |
| MA | Pb | Pb | Cl | 1.15 | |
| MA | Pb | Pb | Br | 0.67 | |
| MA | Pb | Pb | I | 0.27 | |

**Table S5** Computed positions of the valence band maximum (VBM) and conduction band minimum (CBM) for $(MA)_2B^IB^{III}X_6$

| A  | $B^I$ | $B^{III}$ | X  | VBM    | CBM |
|----|-------|-----------|----|--------|-----|
| MA | K     | Bi        | Cl | L      | A   |
| MA | K     | Bi        | Br | Near L | A   |
| MA | K     | Bi        | I  | Γ      | A   |
| MA | Tl    | Bi        | Cl | Γ      | Γ   |
| MA | Tl    | Bi        | Br | Γ      | Γ   |
| MA | Tl    | Bi        | I  | Γ      | Γ   |
| MA | Cu    | Bi        | Cl | M      | L   |
| MA | Cu    | Bi        | Br | M      | L   |
| MA | Cu    | Bi        | I  | M      | L   |
| MA | Ag    | Bi        | Cl | M      | L   |
| MA | Ag    | Bi        | Br | M      | L   |
| MA | Ag    | Bi        | I  | M      | L   |
| MA | Pb    | Pb        | Cl | Γ      | Γ   |
| MA | Pb    | Pb        | Br | Γ      | Γ   |
| MA | Pb    | Pb        | I  | Γ      | Γ   |

**Table S6** Computed single crystal elastic stiffness constants ($C_{ij}$) of $(MA)_2B^IB^{III}X_6$. All units are in GPa

| A | $B^I$ | $B^{III}$ | X | $C_{11}$ | $C_{12}$ | $C_{13}$ | $C_{14}$ | $C_{33}$ | $C_{44}$ | $C_{66}$ |
|---|---|---|---|---|---|---|---|---|---|---|
| MA | K | Bi | Cl | 31.75 | 11.75 | 14.38 | -3.58 | 24.79 | 12.11 | 10.00 |
| MA | K | Bi | Br | 25.03 | 9.13 | 11.99 | -3.89 | 18.90 | 10.68 | 7.95 |
| MA | K | Bi | I | 19.60 | 7.15 | 9.97 | -3.58 | 17.40 | 8.94 | 6.23 |
| MA | Tl | Bi | Cl | 29.79 | 10.73 | 14.28 | -4.29 | 21.45 | 11.86 | 9.53 |
| MA | Tl | Bi | Br | 25.91 | 9.43 | 13.36 | -4.28 | 22.08 | 11.35 | 8.24 |
| MA | Tl | Bi | I | 24.29 | 8.52 | 11.72 | -4.59 | 19.42 | 10.69 | 7.89 |
| MA | Cu | Bi | Cl | 55.40 | 33.67 | 21.95 | 13.81 | 57.62 | 5.70 | 10.86 |
| MA | Cu | Bi | Br | 50.49 | 26.65 | 20.75 | 7.96 | 43.25 | 4.79 | 11.92 |
| MA | Cu | Bi | I | 44.75 | 20.24 | 15.96 | 5.19 | 38.27 | 6.15 | 12.26 |
| MA | Ag | Bi | Cl | 47.23 | 28.00 | 24.07 | 3.44 | 36.46 | 10.30 | 9.61 |
| MA | Ag | Bi | Br | 43.93 | 20.64 | 18.87 | 3.04 | 34.39 | 9.19 | 11.65 |
| MA | Ag | Bi | I | 38.21 | 16.99 | 16.02 | 1.19 | 33.22 | 8.84 | 10.61 |
| MA | Pb | Pb | Cl | 45.67 | 17.40 | 22.15 | -6.56 | 33.22 | 17.91 | 14.14 |
| MA | Pb | Pb | Br | 40.67 | 13.86 | 18.48 | -6.30 | 29.26 | 15.93 | 13.41 |
| MA | Pb | Pb | I | 33.15 | 10.92 | 15.08 | -5.76 | 23.82 | 13.96 | 11.12 |

**Table S7** Computed polycrystalline elastic Young's modulus (E), bulk modulus (B), shear modulus (G), Poisson's ratio ($\nu$) as well as the range of corresponding single crystal elastic moduli of $(MA)_2B^IB^{III}X_6$. All units are in GPa except Poisson's ratio. $(MA)_2CuBiCl_6$ and $(MA)_2CuBiBr_6$ are not mechanically stable therefore values are not shown.

| A | $B^I$ | $B^{III}$ | X | E | B | G | $\nu$ | $E_{max}$ | $E_{min}$ | $G_{max}$ | $G_{min}$ | $\nu_{max}$ | $\nu_{min}$ |
|---|---|---|---|---|---|---|---|---|---|---|---|---|---|
| MA | K | Bi | Cl | 24.03 | 18.75 | 9.34 | 0.29 | 35.17 | 15.29 | 14.78 | 6.28 | 0.50 | 0.10 |
| MA | K | Bi | Br | 18.85 | 14.97 | 7.31 | 0.29 | 31.06 | 10.49 | 13.44 | 4.28 | 0.58 | 0.05 |
| MA | K | Bi | I | 15.33 | 12.31 | 5.93 | 0.29 | 26.14 | 9.97 | 11.41 | 3.66 | 0.62 | 0.06 |
| MA | Tl | Bi | Cl | 21.70 | 17.64 | 8.38 | 0.29 | 35.38 | 11.38 | 15.13 | 4.78 | 0.56 | 0.07 |
| MA | Tl | Bi | Br | 19.90 | 16.24 | 7.68 | 0.30 | 33.26 | 11.98 | 14.35 | 4.59 | 0.60 | 0.07 |
| MA | Tl | Bi | I | 18.42 | 14.64 | 7.14 | 0.29 | 32.01 | 11.05 | 14.09 | 4.24 | 0.60 | 0.06 |
| MA | Cu | Bi | Cl | - | - | - | - | - | - | - | - | - | - |
| MA | Cu | Bi | Br | - | - | - | - | - | - | - | - | - | - |
| MA | Cu | Bi | I | 21.64 | 25.62 | 7.96 | 0.36 | 30.43 | 9.20 | 15.22 | 3.18 | 0.96 | 0.25 |
| MA | Ag | Bi | Cl | 25.08 | 31.00 | 9.19 | 0.37 | 35.45 | 21.06 | 13.41 | 6.50 | 0.51 | 0.19 |
| MA | Ag | Bi | Br | 26.23 | 26.31 | 9.83 | 0.33 | 23.36 | 19.56 | 13.70 | 7.14 | 0.48 | 0.29 |
| MA | Ag | Bi | I | 25.29 | 23.01 | 9.60 | 0.32 | 23.92 | 21.99 | 11.21 | 8.24 | 0.39 | 0.29 |
| MA | Pb | Pb | Cl | 32.26 | 27.39 | 12.37 | 0.30 | 52.53 | 17.66 | 22.63 | 7.28 | 0.56 | 0.07 |
| MA | Pb | Pb | Br | 29.87 | 23.45 | 11.60 | 0.29 | 48.74 | 16.74 | 21.09 | 6.94 | 0.55 | 0.08 |
| MA | Pb | Pb | I | 24.82 | 19.05 | 9.68 | 0.28 | 41.96 | 13.50 | 18.48 | 5.56 | 0.57 | 0.06 |

**Table S8** Computed tolerance factors of $(MA)_2B^IB^{III}X_6$

|    | Cl    | Br    | I     |
|----|-------|-------|-------|
| K  | 0.933 | 0.923 | 0.907 |
| Cu | 1.038 | 1.021 | 0.997 |
| Tl | 0.915 | 0.905 | 0.892 |
| Ag | 0.970 | 0.957 | 0.939 |
| Pb | 0.938 | 0.927 | 0.912 |

**Table S9** Fractional coordinates of $(MA)_2TlBiBr_6$ obtained from single crystal XRD

|     | x         | y    | z    | Uiso       |
|-----|-----------|------|------|------------|
| Bi  | 0         | 0    | 0    | 0.0340(6)  |
| Tl  | 0.5       | 0    | 0    | 0.0560(7)  |
| Br  | 0.2362(4) | 0    | 0    | 0.125(2)   |
| C/N | 0.19      | 0.25 | 0.25 | 0.106(19)  |

**Table S10** Crystal refinement details for the disordered structure model.

| | |
|---|---|
| Empirical formula | $C_2N_2Br_6TlBi$ |
| Formula weight | 950.9 |
| Temperature/K | 298.3(3) |
| Crystal system | cubic |
| Space group | $Fm\bar{3}m$ |
| a/Å | 11.7616(5) |
| b/Å | 11.7616(5) |
| c/Å | 11.7616(5) |
| α/° | 90 |
| β/° | 90 |
| γ/° | 90 |
| Volume/Å$^3$ | 1627.0(2) |
| Z | 4 |
| μ/mm$^{-1}$ | 35.394 |
| F(000) | 1624 |
| Crystal size/mm$^3$ | 0.1 × 0.1 × 0.1 |
| Radiation | MoKα (λ = 0.71073) |
| 2θ range for data collection/° | 6 to 56.198 |
| Index ranges | $-14 \leq h \leq 12$, $-13 \leq k \leq 14$, $-15 \leq l \leq 14$ |
| Reflections collected | 1026 |
| Independent reflections | 132 [$R_{int}$ = 0.0657, $R_{sigma}$ = 0.0287] |
| Data/restraints/parameters | 132/0/9 |
| Goodness-of-fit on $F^2$ | 1.83 |
| Final R indexes [I>=2σ (I)] | $R_1$ = 0.0364, $wR_2$ = 0.0432 |
| Final R indexes [all data] | $R_1$ = 0.0531, $wR_2$ = 0.0449 |
| Largest diff. peak/hole / e Å$^{-3}$ | 1.85/-2.17 |

**Experimental details**

The starting material MABr was prepared by mixing stoichiometric amount of methylamine solution (40wt% in $H_2O$) and HCl (37% in $H_2O$) at 0°C, heated at 60°C to dryness, then washing with acetone and drying overnight in vacuum oven.

Crystal structure determination was carried out using an Oxford Gemini E Ultra diffractometer, Mo Kα radiation (λ = 0.71073Å), equipped with an Eos CCD detector. Data collection and reduction were conducted using CrysAliPro (Agilent Technologies). An empirical absorption correction was applied with the Olex2 platform,[8] and the structure was solved using ShelXT[9] and refined by ShelXL.[10]

8    O. V. Dolomano, L. J. Bourhis, R. J. Gildea, J.A.K. Howard and H. Puschmann, *J. Appl. Cryst*, 2009, **42**, 339-341.

9    G. M. Sheldrick, *Acta Cryst. A*, 2015, **71**, 3-8.

10   G. M. Sheldrick, *Acta Cryst. A*, 2008, **64**, 112-122.